\documentclass[a4paper,11pt]{article}

\usepackage{jheppub} 

\usepackage[utf8x]{inputenc}
\usepackage{amsfonts}
\usepackage{latexsym}
\usepackage{amsmath}
\usepackage{amssymb}
\usepackage{amssymb}

\usepackage{slashed}
\usepackage{upgreek}
\usepackage{mathrsfs}
\usepackage{bbm}

\usepackage{relsize}
\usepackage{graphicx}

\font\mybb=msbm10 at 11pt

\def\bb#1{\hbox{\mybb#1}}

\def\bZ {\bb{Z}}
\def\bR {\bb{R}}

\def\bC {\bb{C}}

\newcommand{\bea}{\begin{eqnarray}}
\newcommand{\eea}{\end{eqnarray}}

\title{\boldmath C-spaces, generalized geometry and double field theory}

\author[]{G. Papadopoulos}

\affiliation[]{ Department of Mathematics\\
King's College London\\
Strand\\
London WC2R 2LS, UK}

\emailAdd{george.papadopoulos@kcl.ac.uk}

\abstract{We construct  a C-space associated with every closed 3-form on a spacetime $M$ and show that it depends  on the
class of the form  in  $H^3(M, \bZ)$. We also demonstrate that C-spaces  have a relation  to  generalized geometry and
 to gerbes.
 C-spaces are constructed after introducing  additional coordinates at the open sets and at their double overlaps of a spacetime generalizing
the standard construction of Kaluza-Klein spaces for 2-forms. C-spaces may not be manifolds and satisfy the topological geometrization condition.
Double spaces arise as local subspaces of C-spaces
that cannot be globally extended. This indicates that for the global definition of double field theories additional coordinates are needed. We explore several other aspect of C-spaces like  their topology and relation to  Whitehead towers, and also describe the construction of C-spaces for closed k-forms.}


\begin{document}
\maketitle
\flushbottom

\section{Introduction}

Double  field theory (DFT) has been introduced to provide a geometric interpretation of the T-duality symmetries and to describe
string theory in a T-duality covariant way, see \cite{duff, arkady, siegel} for early works and \cite{hull1}-\cite{bakas} for more recent developments. More general proposals\footnote{See \cite{nicolai1} for an early work on the geometrization of dualities.} include  $E_{11}$
\cite{west1, west2} and exceptional field theories, see eg  \cite{hull3}-\cite{ grana},  and reviews \cite{rev1, rev2, rev3} and references with.  For the construction of DFT, the spacetime $M$ is enhanced with additional coordinates, leading to a double
 space $D_M$ which has dimension twice that of spacetime. So far the construction of local  actions relies on two ingredients.  First, the use of infinitesimal transformations to prove invariance,
and second the application of the strong section condition.  These infinitesimal transformations combine the spacetime diffeomorphisms and the gauge transformations of the $B$-field  that act on a generalized metric.
This generalized metric is constructed from both the spacetime metric and the $B$-field. This is interpreted as a geometrization of $B$-field. The strong section condition in effect restricts the fields and their infinitesimal transformations to dependent on either the
spacetime or dual coordinates. More recently several suggestions have been made to integrate the infinitesimal transformations of DFT leading to
the construction of finite transformations for the double spaces and for the associated
fields \cite{hz, rev3, perry1, hull3x}. Another suggestion is to employ a non-trivial split metric  on the extended spaces \cite{cederwallx}.
Similar results also hold for the exceptional field theories, however see also \cite{nicolai2, cartan}.

The global definition of DFTs remains an open problem. Using the solution of the strong section condition for the spacetime presented
 in \cite{hz, rev3}, it has been shown in \cite{cfolds} that the patching of double spaces\footnote{Examples of double spaces have been investigated in \cite{cfolds}
from the patching point of view and they have been found to depend on the choice of atlas. Therefore they are not general covariant.}
 constructed  {\it is  consistent if and only if
the 3-form field strength is   exact}. In section 5, we shall  strengthen this statement. The C-spaces that we propose below {\it resolve this global patching problem}.

To identify the spaces which can implement the geometrization of the $B$-field in the context of DFT,
 it has also been argued in  \cite{cfolds} that one necessary ingredient is the {\it topological geometrization condition}. This can be stated as follows:  Given a manifold $M$, eg spacetime, and a closed k-form $\omega^k$, a space\footnote{The spaces that satisfy the topological geometrization property have been called  C-spaces because the topological charge carried by $\omega^k$ is stored in the transition functions of $C$. } $C_M$ satisfies the topological geometrization condition, if and only if there is a projection $\pi:  C_M\rightarrow M$  such that
   $\pi^*\omega^k$ represents the trivial class in $H^k(C_M)$.

    Given $M$ and $\omega^k$, this definition does not uniquely specify $C_M$. There are several constructions of C-spaces via K-theory  and homotopy theory. The latter applies for any manifold and for any form of any degree.
   The standard examples of C-spaces are  circle bundles
   over $M$ which satisfy the topological geometrization property for closed 2-forms, and implement the geometrization of the Maxwell fields in the context of Kaluza-Klein theory.

In this paper, a construction of C-spaces, $C_M^{[\omega^3]}$, is proposed for every closed 3-form, $\omega^3$, on a manifold $M$ provided that $[\omega^3]\in H^3(M, \bZ)$
which is suitable for applications in DFT.
The construction involves the introduction of new coordinates associated with the gauge transformations of the transition functions
of $\omega^3$ at double overlaps with respect to both the \v Cech and de Rham differentials. This leads to  an additional
\begin{itemize}

\item (local) {\it 1-form
coordinate $y^1$  for every open set of spacetime},  as for  double spaces\footnote{However, we shall demonstrate that $y^1$ and the corresponding coordinates for double spaces
transform differently.},

\item and {\it a new angular coordinate $\theta$ at every double intersection of two open sets}.

\end{itemize}
 Exploring the consistency
of the patching conditions given in (\ref{gcspaces}) at triple and 4-fold overlaps, it leads to the requirement that $\omega^3$ must represent a class
in $H^3(M, \bZ)$ as expected from the Dirac quantization condition.  In addition, it is demonstrated that $C_M^{[\omega^3]}$  depends on $[\omega^3]$,
ie it is independent from the choice of a representative of the cohomology class $[\omega^3]$.
From construction is apparent that  $C_M^{[\omega^3]}$ are not manifolds, in particular they may not have a well-defined dimension. Nevertheless they
 can be described in some detail using the transition functions and the additional coordinates. Furthermore, one can show that $C_M^{[\omega^3]}$ satisfies
 the topological geometrization condition.

This construction of C-spaces for closed 3-forms is related to gerbes. In particular, we explain how from $C_M^{[\omega^3]}$ one can construct the gerbe transition functions that arise in the approach of \cite{ghitchin}.
However the construction of $C_M^{[\omega^3]}$ involves the open sets and  their double overlaps, as well as the triple and 4-fold overlaps, in an essential way,   and the emphasis is on the object itself rather than its transition functions on $M$. This is more close in spirit to the definition of gerbes in terms
of sheafs \cite{gbry}  but without the complications of category theory. Furthermore, the construction of $C_M^{[\omega^3]}$ gives a geometric interpretation into   generalized geometry on $M$ as described by Hitchin and Gualtieri \cite{hitchin, gualtieri}.  In particular we shall show that the twisted Courant bracket on the spacetime can be derived from a Courant bracket
on $C_M^{[\omega^3]}$. As result one can define a generalized metric  and carry out generalized differential geometry calculus on $M$.

To get some insight into the topological structure of $C_M^{[\omega^3]}$, we consider the nerve of the good cover of $M$ which provides a chain complex description of $M$.
We find that every  2-simplex in the nerve of $M$ together with the new angular coordinates give rise to a $\bC P^2$ in $C_M^{[\omega^3]}$. We use this to raise the question whether
this construction of  $C_M^{[\omega^3]}$ is related to Whitehead towers. Furthermore, we construct, $C_{T^3}^{[\omega^3]}$, which is the C-space of 3-torus with a 3-form flux.
We demonstrate that $C_{T^3}^{[\omega^3]}$ {\it resolves the patching problems of the double space construction } of \cite{rev3} for this model.

To elucidate the relation between C-spaces and double spaces, we revisit the global properties of the double spaces. We show that
the mere use of the strong section condition, ie without invoking any information about the transformation of the generalized fields,
together with the requirement of the general covariance of the spacetime imply that the double space must be diffeomorphic to $T^*M$. Such a space cannot satisfy the topological geometrization property
and also is in conflict with established examples of T-dual pairs. Moreover if the transition functions of the B-field are related in a linear way to those of the dual coordinates, then the 3-form flux is exact.

Furthermore, we demonstrate that the C-spaces $C_M^{[\omega^3]}$ locally  {\it include} the double spaces. In particular,  the double spaces arise as subspaces of
$C_M^{[\omega^3]}$ after taking  the new angular coordinates $\theta$ to vanish. This can be consistently done only  at appropriate open sets and not globally over the whole
spacetime $M$. Therefore double spaces can only provide a local description  DFTs, ie on a patch of $M$. For the global definition
of DFTs over $M$ additional coordinates are required.

The construction of C-spaces, $C_M^{[\omega^k]}$, can be generalized to every k-form, $\omega^k$, which represents a class in $H^k(M, \bZ)$. This proceeds in a similar way to
that of $C_M^{[\omega^3]}$. However, the construction of $C_M^{[\omega^k]}$ requires the presence of  additional coordinates which are introduced at the multiple intersections
of open sets of $M$.
The properties of  $C_M^{[\omega^k]}$ are also similar, ie $C_M^{[\omega^k]}$ satisfy the topological geometrization condition and depend on the class of $\omega^k$ in $H^k(M, \bZ)$.
  The extended space associated with a k-form, which is the generalization of a double space for $k>3$,  can be seen as a local subspace of $C_M^{[\omega^k]}$. This again indicates that more coordinates are need for the global description of exceptional field theories.

There is a construction of C-spaces  in the context of homotopy theory  using  Whitehead towers. Here we revisit the theory and point out that the Whitehead towers construction
for 2-forms coincides, up to homotopy, with the standard circle bundle construction of Kaluza-Klein spaces. Then we review some of the  properties
of  Whitehead towers construction for closed 3-forms and  ask the question how these are related to $C^{[\omega^3]}_M$ spaces. We also argue that the total space $X_3$ of the Whitehead fibration
$\bC P^\infty \rightarrow X_3\rightarrow S^3$ provides a homotopy description of the gerbe
associated with the generator of $H^3(S^3, \bZ)$.

This paper has been organized as follows. In section 2, we describe the construction of $C_M^{[\omega^3]}$. In section 3, we explain the relation of $C_M^{[\omega^3]}$
to generalized geometry and gerbes. In section 4, we investigate some of the topological properties of $C_M^{[\omega^3]}$ and present the 3-torus with 3-form flux
C-space. In section 5, we explore the applications of C-spaces to DFT. In section 6, we construct C-spaces for closed k-forms. In section 7, we explore the relation
between C-spaces and Whitehead towers, and in section 8, we give our conclusions.

\section{C-spaces for closed 3-forms}

\subsection{C-spaces for closed 2-forms}

Before, we proceed to give the patching conditions of C-spaces associated with closed 3-forms,  let us briefly review the standard Kaluza-Klein space, $C_M^{[\omega^2]}$, for 2-forms.
Let $M$ be a manifold and  $\{U_\alpha\}_{\alpha\in I}$ be a good cover\footnote{Good covers exist for compact and non-compact manifolds and are essential in \v Cech- de Rham theory.} of $M$, for the precise definition see eg \cite{bott} page 42. Moreover suppose that $\omega^2$ represents a class in $H^2(M, \bR)$. Then within the \v Cech-de Rham theory applying the Poincar\'e
lemma on the open sets $U_\alpha$ as well as their
 $U_{\alpha\beta}$ and $U_{\alpha\beta\gamma}$  intersections\footnote{We use the notation $U_{{\alpha_0} \dots {\alpha_k}}=U_{\alpha_0}\cap \dots\cap U_{\alpha_k}$ for the
k-fold intersections or overlaps of open sets.}
\bea
\omega^2=dA^1_\alpha~,~~~-A^1_\alpha+ A^1_\beta=da^0_{\alpha\beta}~,~~~-a^0_{\alpha\beta}-a^0_{\beta\gamma}-a^0_{\gamma\alpha}=2\pi n_{\alpha\beta\gamma}~.
\label{cdr2}
\eea
The Kaluza-Klein space $C_M^{[\omega^2]}$  is constructed from $M$ by introducing a new coordinate $\tau_\alpha$  at each open set $U_\alpha$ with patching conditions
\bea
\big(-\tau_\alpha+\tau_\beta- a^0_{\alpha\beta}\big)~~~ \mathrm{mod}~2\pi \bZ=0~,
\eea
which is consistent at triple overlaps $U_{\alpha\beta\gamma}$ if and only if $n_{\alpha\beta\gamma}\in \bZ$ and so ${1\over2\pi} [\omega^2]\in H^2(M, \bZ)$. Taking the exterior derivative of patching condition, one finds that
$d\tau_\alpha- A^1_\alpha= d\tau_\beta- A^1_\beta$
and so $d\tau-A^1$ is globally defined on the total space $C_M^{[\omega^2]}$. Thus $\pi^*\omega^2= -d(d\tau-A^1)$ is an exact form on $C_M^{[\omega^2]}$, and so  $C_M^{[\omega^2]}$
satisfies the topological geometrization condition. Of course
$C_M^{[\omega^2]}$ is a circle bundle on $M$ with first Chern class given by ${1\over2\pi}[\omega^2]$.

\subsection{Patching C-spaces for closed 3-forms}

To begin the construction of $C_M^{[\omega^3]}$ spaces, suppose $M$ be a manifold and $\omega^3$ be a closed 3-form on $M$.  For  applications in DFT,  $M$
is the spacetime and $\omega^3$ is the NS-NS 3-form field strength. In addition  let $\{U_\alpha\}_{\alpha\in I}$ be a good cover of $M$ as for  2-forms in the previous section. Applying the  Poincar\'e
lemma  on the open sets $U_\alpha$ as well double, triple and 4-fold intersections,
one finds that
\bea
&&\omega^3_\alpha=dB^2_\alpha~,~~~
-B^2_\alpha+B^2_\beta=d a^1_{\alpha\beta}~,~~~-a^1_{\alpha\beta}- a^1_{\beta\gamma}- a^1_{\gamma\alpha}= d a^0_{\alpha\beta\gamma}~,
\cr
&&-a^0_{\beta\gamma\delta}+a^0_{\alpha\gamma\delta}-a^0_{\alpha\beta\delta}+ a^0_{\alpha\beta\gamma}= 2\pi n_{\alpha\beta\gamma\delta}~,
\label{kformdata}
\eea
respectively, where $n_{\alpha\beta\gamma\delta}$ are constants and the combinatorics of the open set labels follow from the definition of the \v Cech differential, see (\ref{cechd}). $B_\alpha$ are the 2-form gauge potentials of $\omega^3$ on each $U_\alpha$, and  $\{a^1_{\alpha\beta}, a^0_{\alpha\beta\gamma}\}$ are the patching
or transition ``functions'' of $\omega^3$ at double and triple
overlaps.  Moreover if ${1\over 2\pi}\omega^3$ represents a class in $H^3(M, \bZ)$, then  $n_{\alpha\beta\gamma\delta}\in \bZ$ on all 4-fold overlaps, $U_{\alpha\beta\gamma\delta}$.
All the patching data are skew-symmetric under the exchange of open set labels, ie $a^1_{\alpha\beta}=-a^1_{\beta\alpha}$ and similarly for the rest.

The  gauge potentials $B_\alpha$ and the transition functions $\{a^1_{\alpha\beta}, a^0_{\alpha\beta\gamma}\}$  are not uniquely defined. In fact, the gauge potentials are defined up to the gauge transformations
\bea
B'_\alpha=B_\alpha+ d\zeta_\alpha^1~,
\label{gttran0}
\eea
and similarly the transition functions are defined up to gauge transformations as
\bea
a'^1_{\alpha\beta}&=&a^1_{\alpha\beta}-\zeta_\alpha^1+ \zeta_\beta^1+ d \zeta^0_{\alpha\beta}~,
\cr
a'^0_{\alpha\beta\gamma}&=&a^0_{\alpha\beta\gamma}-\zeta^0_{\alpha\beta}- \zeta^0_{\beta\gamma}- \zeta^0_{\gamma\alpha}~.
\label{gttran}
\eea
These gauge transformations are the only ones compatible with the closure of $\omega^3$.

The construction of $C^{[\omega^3]}_M$ proceeds with the introduction of new coordinates $y^1_\alpha$ and $\theta_{\alpha\beta}$  associated with the open sets $U_\alpha$ and
the double overlaps $U_{\alpha\beta}$, respectively. These are new coordinates in addition to those of the spacetime. They should be thought in the same way  as the Kaluza-Klein coordinate $\tau$ that we have introduced for the description of $C^{[\omega^2]}_M$ in the previous section. Though $y^1$ is assigned
 the degree of a 1-form. In addition, one imposes the  patching  conditions
\bea
-y^1_\alpha+y^1_\beta+d\theta_{\alpha\beta}&=& a^1_{\alpha\beta}~,
\cr
\big (\theta_{\alpha\beta}+ \theta_{\beta\gamma}+ \theta_{\gamma\alpha}+  a^0_{\alpha\beta\gamma}\big)&=&0 ~~~\mathrm{mod}~ 2\pi \bZ~,
\label{gcspaces}
\eea
on $U_{\alpha\beta}$ and $U_{\alpha\beta\gamma}$.

Using the second condition in (\ref{kformdata}), one finds that consistency of the first condition on triple overlaps yields
\bea
d(\theta_{\alpha\beta}+\theta_{\beta\gamma}+\theta_{\gamma\alpha}+ a^0_{\alpha\beta\gamma})=0~.
\label{ffeqn}
\eea
 This is implied from the second  condition in (\ref{gcspaces}). Next investigating the consistency
of the second condition of (\ref{gcspaces}) on  4-fold overlaps and after using the last condition in (\ref{kformdata}), one finds that
\bea
 n_{\alpha\beta\gamma\delta}=0 ~~~\mathrm{mod}~ \bZ~.
\eea
This is satisfied provided that ${1\over2\pi}\omega^3$ represents a class in $H^3(M, \bZ)$.

One of the questions that arises in imposing (\ref{gcspaces}) is  how  one is supposed to think about these new coordinates and their patching conditions.
The coordinates should be thought in the same way as in the usual construction of a circle bundle over a manifold utilizing the patching conditions
of a manifold together with those of a closed 2-form. For the gluing transformations,
this particularly applies to the second patching condition which involves triple overlaps and three coordinates rather than double overlaps and two coordinates which appear in
the usual patching  for manifolds.
To give some insight into this question, one can view the usual patching of manifolds as follows.  Given two charts, ie open sets and coordinates adapted to each one of the sets,
the patching condition at the double intersection relates the coordinates of first chart to the coordinates of the second chart, and vice versa.  In this context,  the second patching condition
in (\ref{gcspaces}) specifies how the three $\theta$ coordinates, each one
associated with one of the three double overlaps that contribute to the triple overlap, are related.

\subsection{The $y^1$ coordinates}

To get some insight into the nature of $y$ coordinates, observe that the first patching condition in (\ref{gcspaces}) can be solved to express
the $y^1$ coordinates in terms of the angular coordinates $\theta$ using a partition of unity $\{\rho_\alpha\}_{\alpha\in I}$ subordinate to $\{U_\alpha\}_{\alpha\in I}$; for the definition of partitions of unity see eg \cite{bott} page 21.  In particular, one has that
\bea
y^1_\alpha=\tilde y^1_\alpha+ \sum_\gamma \rho_\gamma (d\theta_{\alpha\gamma}-a^1_{\alpha\gamma})~,
\label{ycor}
\eea
where $\tilde y^1_\alpha$ are coordinates which transform as   1-forms on $M$, $\tilde y^1_\alpha=\tilde y^1_\beta$. So  $y_\alpha^1$ are coordinates
 which transform as 1-forms of $M$ on $U_\alpha$ and receive
an additional correction from the angular coordinates $\theta_{\alpha\gamma}$ and the transition functions $a^1_{\alpha\gamma}$ when they approach the double overlaps $U_{\alpha\gamma}$.

One of the consequences of (\ref{ycor}) is that the C-spaces $C_M^{[\omega^3]}$  are not manifolds. To see this, first observe that  by construction there is a projection
$\pi:~~ C_M^{[\omega^3]}\rightarrow M$.  The dimension of the inverse image $\pi^{-1}(x)$ of $x\in M$  depends on $x$. If $x\in U_\alpha$ and $x\notin U_{{\alpha_0} \dots {\alpha_k}}$,
$\pi^{-1}(x)=\bR^n$. While if $x\in U_{\alpha\beta}$ and $x\notin U_{\alpha\beta\gamma}$, then $\pi^{-1}(x)=\bR^n\times S^1$. Finally if $x\in U_{\alpha\beta\gamma}$, then $\pi^{-1}(x)=
\bR^n\times T^2$, and so on. As a consequence $C_M^{[\omega^3]}$ may not have a well-defined dimension.

\subsection{Dependence on $\omega^3$}

Here we shall investigate whether or not  $C_M^{[\omega^3]}$  depends on the representative $\omega^3$ of the class ${1\over2\pi}[\omega^3]\in H^3(M,\bZ)$.
Suppose that $\omega'^3$ is another representative of  $[\omega^3]$, ie $ [\omega'^3] =[\omega^3]$.  Then there is a globally defined 2-form $u^2$
such that $\omega'^3=\omega^3+du^2$. Thus $B'^2_\alpha=B^2_\alpha+ u^2_\alpha$.  Since $u^2_\alpha=u^2_\beta$ at double overlaps the dependence on $u$
drops out and so $a^1_{\alpha\beta}$ does not dependent on the choice of representative of $[\omega^3]$. As a consequence the transition functions
of $C_M^{[\omega^3]}$ do not depend on the representative of $[\omega^3]$.

There is  additional gauge redundancy  in the definition of  $B_\alpha$ and that of the transition
functions given in (\ref{gttran0}) and (\ref{gttran}), respectively. This is eliminated
by performing the compensating transformations
\bea
y'^1_\alpha=y^1_\alpha+ \zeta_\alpha^1~,~~~~\theta'_{\alpha\beta}=\theta_{\alpha\beta}+\zeta^0_{\alpha\beta}~,
\eea
on the new coordinates. As a result  $C_M^{[\omega^3]}$ does not depend on the choices made including that of the representative of $[\omega^3]$.

\subsection{Dependence on the cover}

One should also investigate a more subtle choice in the construction of  $C_M^{[\omega^3]}$ that of the good cover $\{U_\alpha\}_{\alpha\in I}$. For this one can adapt
a similar strategy as the one for manifolds which one starts from an atlas and adds all the compatible
charts, ie all the charts which have smooth transition functions for each intersection amongst themselves and each intersection with the charts of the original atlas.  Such a construction leads to the notion of the maximal atlas which characterizes the smooth structure on a manifold. Therefore if one begins with a topological space that can be given a manifold structure with respect to two different atlases
which however lead to the same maximal atlas, then the two original spaces are identified as manifolds, ie the two original atlases give rise to the same smooth structure on the topological space.   There is not necessarily a unique maximal atlas on a manifold as it is known that on a given topological manifold there can be more than one  smooth structures.

In the same way one can add to a good cover $\{U_\alpha\}_{\alpha\in I}$ all the additional open sets
(charts) which are compatible with the smooth structure of $M$ and give rise to a new maximal good cover on $M$. Then with respect to this maximal good cover one can define  $C_M^{[\omega^3]}$. Moreover, one can assert that if two good covers give rise to the same maximal good cover, then the two original C-spaces must be identified. It is not apparent  how the C-spaces with respect to two different maximal good covers are related. There can be a moduli of possibilities but this will not be unusual as
many constructions and structures on spaces depend on the choice of open covers and atlases\footnote{The notion of the smooth structure is indeed atlas dependent. For example if one considers a triangle and takes the atlas induced on it as a $\bR^2$ subspace, then the triangle is not a manifold because of the cusp singularities. However, the triangle
is homeomorphic to a circle and so there is another atlas on the triangle inducing on it a smooth structure. The new atlas can be constructed explicitly. }. Moreover the existence of a moduli will not invalidate the construction as
each space provides a solution to the patching problem which has been the main question that has been addressed in this paper. It will simply mean that we have more than one solutions and the implications of this will be of interest to investigate. Furthermore, there is a mild indication that all such C-spaces have the same homotopy type as all of them have the effect to trivialize the
class of $\omega^3$ and cohomology groups are homotopic invariant.

\subsection{Topological geometrization condition}

It has been argued in \cite{cfolds} that any space which geometrizes a k-form flux must be a C-space, ie it  admits a projection onto
the spacetime such that the pull back of the k-form flux represents the trivial cohomological class in the C-space.

Here we shall demonstrate that $C_M^{[\omega^3]}$ is a C-space. As we have mentioned, there is a projection $\pi$ from $C_M^{[\omega^3]}$ onto $M$. Next taking the differential of the first patching condition in (\ref{gcspaces}), one finds that
\bea
-dy^1_\alpha+dy^1_\beta=d a^1_{\alpha\beta}~.
\eea
Using the second condition in (\ref{kformdata}), this can be rewritten as
\bea
dy^1_\alpha-B^2_\alpha=dy^1_\beta- B^2_\beta~.
\eea
Therefore $dy^1-B^2$ is globally defined on $C_M^{[\omega^3]}$. As $\pi^*\omega^3=- d(dy^1-B^2)$,  $\pi^*\omega^3$ is exact on $C_M^{[\omega^3]}$. Therefore $C_M^{[\omega^3]}$ satisfies
the topological geometrization property.

\section{Relation to gerbes and generalized geometry}

\subsection{Gerbes}

In the definition of \cite{ghitchin}, a gerbe is the object which represents a class in $H^3(M, \bZ)$ in the same way that a circle  bundle represents a
class in $H^2(M, \bZ)$. It is expected that given a manifold $M$ and a class in $H^3(M, \bZ)$, in a certain sense,  the gerbe is uniquely specified.
In a direct analogy with circle bundles, gerbes  are investigated via their transition functions.  To relate the transition functions of a gerbe as defined in \cite{ghitchin} to the transition functions we use here for $C_M^{[\omega^3]}$, write
\bea
g_{\alpha\beta\gamma}=e^{i a^0_{\alpha\beta\gamma}}~.
\eea
Then the second equation  in (\ref{kformdata}) reads as
\bea
g_{\beta\gamma\delta}^{-1} g_{\alpha\gamma\delta} g^{-1}_{\alpha\beta\delta} g_{\alpha\beta\gamma}=1~,
\eea
which can be recognized as the patching condition of a gerbe on a 4-fold overlap.

Therefore $C_M^{[\omega^3]}$ is a gerbe. But the emphasis in the construction  of $C_M^{[\omega^3]}$ is different. Instead of focusing on the transition
functions,  $C_M^{[\omega^3]}$ describes the object itself.  Furthermore $C_M^{[\omega^3]}$ is possibly one of the many representatives of $[\omega^3]\in H^3(M, \bZ)$
that has been chosen such that it can apply to DFT.
In fact, this is the case even
for the elements of $H^2(M, \bZ)$. To see this note that  these can be represented with complex line bundles $L$ as well. Furthermore $L$ and the direct sum $L\oplus I$, where $I$ is   the trivial $I$ line bundle,  represent the same class  in $H^2(M, \bZ)$.
Clearly $L$ and $L\oplus I$   have different geometric properties which can be essential in certain applications.

The construction of $C_M^{[\omega^3]}$ via the introduction of $y^1$ and $\theta$ coordinates
at the open sets and double overlaps and their patching according to (\ref{kformdata}) are  essential for the applications considered here. Note for example
that for the description of the gerbe patching data $g_{\alpha\beta\gamma}$, these coordinates are not necessary. Presumably
there are other spaces  with different geometric properties from $C_M^{[\omega^3]}$  that represent the same class in $H^3(M, \bZ)$.

\subsection{Generalized geometry}

\subsubsection{Closed 2-forms}

Before we investigate the relation between  $C_M^{[\omega^3]}$ spaces and generalized geometry \cite{hitchin, gualtieri}, it is instructive to examine how the geometry of the
spacetime is related to that of the Kaluza-Klein space $C_M^{[\omega^2]}$. On the spacetime, one can define an extension $E$ of the tangent bundle $TM$ with
a trivial bundle $I$, $0\rightarrow I\xrightarrow{i} E\xrightarrow{j} TM\rightarrow 0$  which has transition functions
\bea
X_\alpha=X_\beta~,~~~f_\alpha=f_\beta-X_\beta (a^0_{\alpha\beta})~,
\label{pat2x}
\eea
where $a^0_{\alpha\beta}$ are the transition functions of the 2-form $\omega^2$.

Choosing a splitting $h: TM\rightarrow E$, one can define
 the twisted bracket given by
 \bea
[h(X)+i(f), h(Y)+i(g)]_{\omega^2}=h([X,Y])+i\big(X(g)-Y(f) +\omega^2(X,Y)\big)~.
\label{twist2}
\eea
This bracket by construction is preserved by the patching conditions (\ref{pat2x}).

To  give a geometric interpretation to the construction above, observe that $TC_M^{[\omega^2]}$ is also an extension of $\pi^*TM$ with a trivial bundle $I$; at every point $p\in C_M^{[\omega^2]}$,   $TC_M^{[\omega^2]}$ has a preferred direction that of the  the tangent bundle
of the fibre $S^1$.  Furthermore $C_M^{[\omega^2]}$ is equipped with a globally defined  1-form $d\tau-A$, which is a principal bundle connection,
 and so  splits  $TC_M^{[\omega^2]}$ into horizontal
and vertical subspaces. In particular the horizontal lift of a vector field $X$ on $TM$ to $TC_M^{[\omega^2]}$ is
\bea
X^h= X^i({\partial\over\partial x^i}+ A_i {\partial\over\partial \tau})~.
\eea
Then observe that  the Lie bracket of the $S^1$-invariant sections of $TC_M^{[\omega^2]}$ which can be written as $X^h+ f {\partial\over\partial \tau}$ is
\bea
[X^h+f{\partial\over\partial \tau}, Y^h+g{\partial\over\partial \tau}]=[X,Y]^h+\big(X(g)-Y(f) +\omega^2(X,Y)\big){\partial\over\partial \tau}~.
\label{kkpart}
\eea
Therefore, the Lie bracket reproduces the twisted bracket (\ref{twist2}) upon setting $h(X)=X^h$ and $i(f)= f{\partial\over\partial \tau}$. From the physics point of
view, the bracket (\ref{kkpart}) arises in the quantization of a charged particle in a magnetic field carrying Kaluza-Klein momentum in the extra direction $\tau$.

\subsubsection{Closed 3-forms}

Generalized geometry \cite{hitchin, gualtieri} on $M$ is based on the extension $E$ of the tangent bundle $TM$,  $0\rightarrow T^*M\xrightarrow{i} E\xrightarrow{j} TM\rightarrow 0$.
The patching conditions of  $E$  are
\bea
X_\alpha=X_\beta~,~~~\zeta_\alpha=\zeta_\beta-\iota_{X_{\beta}} d a^1_{\alpha\beta}~,
\label{patchgg}
\eea
where $da^1_{\alpha\beta}$  satisfies the \v Cech co-cycle condition, $da^1_{\alpha\beta}+
da^1_{\beta\gamma}+ da^1_{\gamma\alpha}=0$, as it can be seen from (\ref{kformdata}).
Choosing a splitting $h: TM\rightarrow E$, one can define
a twisted Courant bracket on $E$ given by
\bea
[h(X)+i(\zeta), h(Y)+i(\eta)]_{\omega^3}^C=h([X,Y])+i\big({\cal L}_X\eta-{\cal L}_Y\zeta-{1\over2} d\big(\eta(X)-\zeta(Y)\big)- \iota_X \iota_Y \omega^3\big)~.
\label{tcb}
\eea
This bracket by construction is well-defined. For use later, the untwisted Courant bracket $[X+\zeta, Y+\eta]^C$, where $X,Y$ are vector fields and $\zeta,\eta$ are 1-forms,
 is defined as above after suppressing the maps $i, h$ and removing the term $\iota_X \iota_Y \omega^3$.

To give a geometric interpretation to the above construction, let us view it from the perspective of
 $C_M^{[\omega^3]}$.  The tangent bundle of $C_M^{[\omega^3]}$
is not well-defined. However, we can define a bundle ${\cal E}$ spanned by  $({\partial\over\partial x^i}, {\partial\over\partial y_j})$, where we note
from (\ref{ycor}) that $dy= dy_i\wedge dx^i$ and ${\partial\over\partial y_j}$ is defined as the dual of $dy_i$, ie $\langle dy_i, {\partial\over\partial y_j}\rangle=\delta^j{}_i$.  We have suppressed the degree  of  $y$ as well as
the open set labeling.  This definition is in direct analogy to that of $TC_M^{[\omega^2]}$ for which the fibres are spanned by
$({\partial\over\partial x^i}, {\partial\over\partial \tau})$,  and  the observation that the patching conditions
\bea
-dy^1_\alpha+dy^1_\beta=da^1_{\alpha\beta}~,
\label{pgg}
\eea
do not depend on the $\theta$ coordinates.   It is clear from the patching conditions that ${\cal E}$ is an extension  $0\rightarrow \pi^*T^*M\xrightarrow{k} {\cal E} \xrightarrow{\ell} \pi^*TM\rightarrow 0$.

To continue observe that $dy-B$ defines a map from $\pi^* TM$ into ${\cal E}^*$. Therefore its dual $(dy-B)^*$ defines a map from ${\cal E}$ to $\pi^* T^*M$ and  this is a analogous structure to a principal bundle connection. So ${\cal E}$ can be split into horizontal and vertical subspaces. In particular
the horizontal lift of a vector field $X$ on $M$ is
\bea
X^h=X^i\big({\partial\over\partial x^i}+B_{ij} {\partial\over\partial y_j}\big)~,
\eea
while the vertical subspace is spanned by $({\partial\over\partial y_j})$.
The sections of ${\cal E}$ which depend only on the coordinates of $M$ can be written as $X^h+k(\zeta)$, where
\bea
k(\zeta)=\zeta_i {\partial\over\partial y_i}~.
\eea
Then observe that the computation of the (untwisted) Courant bracket gives
\bea
[X^h+k(\zeta), Y^h+k(\eta)]^C=[X,Y]^h+k\big({\cal L}_X\eta-{\cal L}_Y\zeta-{1\over2} d\big(\eta(X)-\zeta(Y)\big)- \iota_X \iota_Y \omega^3\big)~.
\eea
The right hand side  gives the twisted Courant bracket (\ref{tcb}) upon setting $h(X)=X^h$ and $i(\zeta)=\zeta_i {\partial\over\partial y_i}$.
Observe that, unlike the naive Lie bracket on ${\cal E}$,  the Courant bracket transforms covariantly under shifts related to the choice of a representative for $B$. This can readily be
 seen from the  expression above.

We can also  globally define tensors on $C_M^{[\omega^3]}$, like for example a generalized metric
\bea
G=g_{ij} dx^i dx^j+ g^{ij} (dy_i+B_{ik} dx^k) (dy_j+B_{j\ell} dx^\ell)~,
\eea
where $g$ is a metric on $M$.

\section{Some topological aspects of $C_M^{[\omega^3]}$ and an example}

\subsection{Topological aspects}

One way to get an insight into the topological structure of a C-space it is instructive to investigate  $C_M^{[\omega^3]}$ in a chain complex approximation of the spacetime.
Given a good cover $\{U_\alpha\}_{\alpha\in I}$ on $M$, one can associate a chain complex with $M$ the nerve ${\cal N}$ of the cover, see eg \cite{bott} page 100.
${\cal N}$  is constructed as follows. One introduces a vertex
for each open set $U_\alpha$ of the cover. Two vertices are joined by a edge if and only if the corresponding open sets intersect $U_{\alpha\beta}\not=\emptyset$. The faces of
three edges of a triangle  are filled if and only if  the corresponding three open sets have a common intersection $U_{\alpha\beta\gamma}\not=\emptyset$, and so on. The cohomology of this chain complex
is exactly the same as the de Rham cohomology or singular cohomology depending on the coefficients.

Let us now focus how the information from the additional coordinates of $C_M^{[\omega^3]}$ can be stored on the nerve   ${\cal N}$. This particularly applies
to the angular coordinates $\theta_{\alpha\beta}$ as the $y^1_\alpha$ coordinates are contractible. It is apparent from the construction of $C_M^{[\omega^3]}$
that the vertices of ${\cal N}$ do not alter as there are no angular coordinates associated to open sets. However a circle is associated to every point
of an edge in   ${\cal N}$ as these represent the intersection of two open sets.  Furthermore at every point on a  face of ${\cal N}$ one should associate
a 2-torus. This is because of the second patching condition in (\ref{gcspaces}) as the three angular coordinates associated to each edge are restricted to two.

Therefore one can describe this construction at a face of  ${\cal N}$ as follows. The 2-torus of the face degenerates to  circles at each of the three edges, and in turn, the circles
at the edges and the tori of the face degenerate to a point as they approach the vertices. Such a structure is reminiscent\footnote{This construction has been adapted
 to construct the universal bundle classifying spaces for any group, see eg  \cite{huse}.} to that of $\bC P^2$.  To see this consider the algebraic equation
of  $S^5$,
\bea
w_1 \bar w_1+w_2 \bar w_2+ w_3 \bar w_3=1~.
\eea
Setting $t_1=w_1 \bar w_1, t_2=w_2 \bar w_2$ and $t_3= w_3 \bar w_3$, this can be seen as the defining equation of a 2-simplex.  The three phases of the
complex numbers $w_1, w_2$ and $w_3$ associate a circle at every vertex, a 2-torus at every point of a edge, and a 3-torus at every point of the face. As $\bC P^2$ is
the base space of the fibration, $S^1\rightarrow S^5\rightarrow \bC P^2$, where $S^1$ acts  from the right on the triplet $(w_1, w_2, w_3)$,
a circle is removed from every point of the simplex leading to the picture describe above for ${\cal N}$.  If such a topology is put on  $C_M^{[\omega^3]}$, it would be
different from that of spacetime $M$.  As we shall see $\bC P^2$, or rather $\bC P^\infty$, appears also in the homotopy approach to C-spaces using
Whitehead towers.

\subsection{The C-space of 3-torus with a 3-form flux}

The construction of $C_M^{[\omega^3]}$ described in section 2 is general and applies to every manifold with a good cover equipped with a closed 3-form which
represents a class in $H^3(M, \bZ)$. As  good covers exist on manifolds, one can construct $C_M^{[\omega^3]}$ for all smooth solutions of supergravity theories
including that of the NS5-brane\footnote{The dilaton singularity does not affect the construction.}.

 Here we construct the C-space of a 3-torus with a 3-form flux. This example was
initially investigated from the perspective of double spaces in \cite{rev3}.  Later it was explored from the patching point of view in \cite{cfolds} where it was found
that the construction depends on the choice of the atlas on $T^3$. Another feature of the construction was that a quantization condition was imposed at the triple
overlaps rather than the 4-fold overlaps as required by the Dirac quantization condition of  3-forms field strengths.

We shall follow the notation of \cite{cfolds} where all the data regarding the patching conditions of the 3-form flux can be found\footnote{Strictly speaking one
should introduce a third open set on $S^1$, $U_3=(-{\pi\over 4}, {\pi\over4})$, so that the cover is a good cover. As the transition functions between $U_1$ and $U_3$, and $U_2$ and $U_3$ are the identity,
there is no change in the computations on \cite{cfolds} and the effects of $U_3$ have already been taken into account via the choice of
$n_x$.}. The patching conditions of the C-space are
\bea
-y^1_{\alpha_1}+y^1_{\alpha_2}+d\theta_{\alpha\beta}&=& a^1_{\alpha_1\alpha_2}~,
\cr
\big (\theta_{\alpha_1\alpha_2}+ \theta_{\alpha_2\alpha_3}+ \theta_{\alpha_3\alpha_1}+  a^0_{\alpha_1\alpha_2\alpha_3}\big)&=&0 ~~~\mathrm{mod}~ 2\pi \bZ~,
\label{tcfold}
\eea
where we have set $\alpha_1=i_1j_1k_1$ and so on. In the atlas we have chosen on $T^3$, the components of $a^1_{\alpha_1\alpha_2}$ and $a^0_{\alpha_1\alpha_2\alpha_3}$ are linear in the coordinates
of $T^3$. However the above patching conditions  do not depend on this choice. This particularly applies to the second condition in (\ref{tcfold}) as the consistency required
for it leads to  $n_{\alpha_1\alpha_2\alpha_3\alpha_4}\in \bZ$ on 4-fold overlaps.  Since $n_{\alpha_1\alpha_2\alpha_3\alpha_4}$ are constant for any choice
of an atlas, the quantization condition is atlas independent. This should be contrasted with the DFT calculation
 which for consistency requires that the components of the 1-form $d a^0_{\alpha_1\alpha_2\alpha_3}$ should be constant and that they should identified periodically  up to some period. As $d a^0_{\alpha_1\alpha_2\alpha_3}$ is a local 1-form, the constancy of its components is an atlas dependent statement
\cite{cfolds}.

\section{DFT on double manifolds}

\subsection{Revisiting the patching of double manifolds}
In the formulation of DFT so far, one introduces a new set of coordinates\footnote{In \cite{cfolds} the dual coordinates were denoted with $y$.  Here we denote them with $\tilde z$
 to distinguish them from those of the C-spaces as they have different transformation properties.} $\tilde x$ in addition to those of the spacetime $x$ and imposes
on all the fields and their transformations the strong section condition which reads
\bea
{\partial\over\partial x^i} F {\partial\over\partial \tilde x_i}G+{\partial\over\partial x^i} G {\partial\over\partial \tilde x_i}F=0~,~~~{\partial\over\partial x^i}{\partial\over\partial \tilde x_i} F=0~.
\label{ssc}
\eea
Setting for $F$ and $G$ the infinitesimal local transformations $\delta x^i$ and $\delta \tilde x_i$ of $x^i$ and $\tilde x_i$, respectively, and assuming that $\delta x^i$ must be arbitrary functions of $x$, which is required
in order to account for all reparameterizations the spacetime\footnote{It is required for example for the construction of a maximal atlas on the spacetime or equivalently general covariance.}, one concludes that the most general solutions to the above conditions are
\bea
\delta x^i=\xi^i(x)~,~~~\delta \tilde x_i= \kappa_i(x)~.
\label{solssc}
\eea
In particular, the first equation in (\ref{ssc}) for $F=G=\delta x^i$ implies that $\delta x^i$ can depend only on $x$. Then again the first equation  for $F=\delta x^i$ and $G=\delta \tilde x_i$ implies that $\delta \tilde x_i$
can dependent only on $x$ as well.
These infinitesimal transformations
can be integrated to give
\bea
x'^i=x'^i(x^j)~,~~~\tilde x'_i=\tilde x_i-\kappa_i(x)~.
\eea
Moreover in \cite{hz, rev3}, $\kappa$ is related linearly to the gauge transformations of the $B$ field. To investigate the global properties of DFTs,  these transformations are interpreted
as patching conditions,
\bea
x_\alpha^i= x^i_{\alpha\beta} (x_\beta)~,~~~\tilde x_\alpha=\tilde x_\beta- \kappa_{\alpha\beta}(x_\beta)~,
\label{dftp}
\eea
where we have introduced a good cover $\{U_\alpha\}_{\alpha\in I}$ on the spacetime $M$.

The strong section condition has another solution where $z$ and $x$ exchange places, this is the solution for the dual space.
It also has many more solutions\footnote{One can easily construct many power series solutions.}
provided that one weakens the requirement that $\delta x^i$ must be an arbitrary function of $x$ and does not allow for general reparametrizations of $M$.  But this breaks general covariance.

So in order to allow for spacetime reparametrization invariance, one is forced to patch the theory with transformations of the type (\ref{dftp}).
If this is the case, then
\bea
\kappa_{\alpha\beta}+\kappa_{\beta\gamma}+\kappa_{\gamma\alpha}=0~.
\label{callc}
\eea
Using the results of \cite{cfolds}, one concludes that this is possible if and only if the double space is diffeomorphic to $D_M=T^*M$. This is because the condition (\ref{callc}) implies that $\kappa_{\alpha\beta}=-\zeta_\alpha+ \zeta_\beta$ and so after a redefinition of the $z$ coordinates transform as 1-forms.

This result is independent from the form of finite transformations on the fields  and  other geometric considerations.
It is a consequence of the application of the strong section condition. {\it Thus if one uses the strong section condition  to describe the double theory and allows for general
reparameterizations of
 the spacetime coordinates, then one is led to the conclusion that  the double space  is   $T^*M$.}

This has immediate consequences. First if the transformations of the dual coordinates $z$ do not transform under the B-field gauge transformation, it appears to contradict standard T-duality results like that of the $S^3$
 and $S^3/\bZ_p$ pair. Consistency of the construction of this T-duality pair requires that the Hopf fibre coordinate $\tilde\theta$  of $S^3/\bZ_p$, which can be identified as the dual coordinate of the fibre
coordinate $\theta$ of $S^3$,  transforms non-trivially under the B-field patching conditions of the $S^3$ solution. {\it Thus the identification of the double space with $T^*M$ is in conflict with examples.}

Furthermore,  $T^*M$ is contractible to $M$, so $\pi^*\omega^3$ is not trivial in $T^*M$. Thus  this space   does not satisfy the
topological geometrization condition. In addition, if the transition  functions of $\omega^3$ at double overlaps are related via a linear transformations
to $\kappa$, then $\omega^3$ is exact \cite{cfolds}.

As a final remark, one can try to patch the double space using both spacetime   and dual space patching conditions as
\bea
x_\alpha= x_{\alpha\beta} (x_\beta)~,~~~\tilde x_\alpha=\tilde x_\beta- \kappa_{\alpha\beta}(x_\beta)~,
\cr
\tilde x_\gamma=\tilde x_{\gamma\alpha}(\tilde x_\alpha)~,~~~x_\gamma=x_\alpha-\tilde\kappa_{\gamma\alpha}(\tilde x_\alpha)
\eea
on $U_{\alpha\beta}$ and $U_{\gamma\alpha}$, respectively, and seek consistency at the triple overlap $U_{\alpha\beta\gamma}$.
However, it is straightforward to see that the patching conditions on $U_{\beta\gamma}$ do not satisfy the
strong section condition.

\subsection{Relation of double spaces to C-spaces}

Now let us compare the results of the previous section with those we have obtained for the $C^{[\omega^3]}_M$ spaces in section 2. In particular, let us compare the second patching condition of (\ref{dftp}) with the first
patching condition in (\ref{gcspaces}). It is clear (\ref{gcspaces}) reduces to (\ref{dftp}) only when the new coordinate $\theta_{\alpha\beta}$ is chosen\footnote{If $\theta_{\alpha\beta}$
was not identified $\mathrm{mod} 2\pi \bZ$, it would have been sufficient to choose it as a function of  $U_{\alpha\beta}$.} as
\bea
\theta_{\alpha\beta}=0~,
\eea
$y^1=z$ and $\kappa_{\alpha\beta}=a^1_{\alpha\beta}$. This choice cannot be  made everywhere on $M$ consistent with the data.
Thus the double spaces are local subspaces of  $C^{[\omega^3]}_M$.

Although the geometric aspects of DFT on $C^{[\omega^3]}_M$ have not been developed, it is clear from the topological considerations presented that {\it for the global definition
of DFT additional coordinates are required}. The mere introduction of $z$ coordinates in the context of double spaces is not sufficient to geometrize the topological charges of $\omega^3$,
and to give  a global definition of double spaces. The examination of the example of \cite{rev3} from the patching point of view in \cite{cfolds}
and in section 4.2 supports this assertion. However, it is not apparent how the additional coordinates $\theta$ can be inserted in the description of DFTs.

\section{C-spaces for closed k-forms}

\subsection{The construction of $C^{[\omega^k]}_M$}

The construction of C-spaces for $\omega^k$ closed forms, $C^{[\omega^k]}_M$,  can be done in a way similar to that for $C^{[\omega^3]}_M$. To simplify
the discussion it is convenient to introduce the \v Cech differential $\delta$.  As before we choose a good cover $\{U_\alpha\}_{\alpha\in I}$ on $M$ and
define
\bea
\delta \lambda^m_{\alpha_0\alpha_1\dots \alpha_p}=\sum_{i=0}^p (-1)^i\lambda^m_{\alpha_0\dots\alpha_{i-1} \hat \alpha_i \alpha_{i+1}\dots \dots \alpha_p}~,
\label{cechd}
\eea
where $\lambda^m$ is a m-form defined at p-overlaps and  restricted upon applying $\delta$ to $(p+1)$-overlaps, and  $\hat \alpha_i$ means that the label $\alpha_i$ is omitted.
As before all these forms defined at the various overlaps are skew-symmetric under the exchange of the labels of the open sets. For example,
\bea
\delta \lambda^m_{\alpha_0\alpha_1}=-\lambda^m_{\alpha_0}+ \lambda^m_{\alpha_1}~,
\eea
on $U_{\alpha_0\alpha_1}$. Observe that $\delta^2=0$ and $d\delta=\delta d$.

Applying the Poincar\'e lemma, the  \v Cech-de Rham expansion of a k-form at multiple overlaps is
\bea
\omega_\alpha^k= dA^{k-1}_\alpha~,~~~\delta A^{k-1}_{\alpha_0\alpha_1}=d a^{k-2}_{\alpha_0\alpha_1}~, \dots~, \delta a^{k-\ell}_{\alpha_0\dots\alpha_\ell}= d a^{k-\ell-1}_{\alpha_0\dots\alpha_\ell}~, \dots~, \delta a^0_{\alpha_0\dots\alpha_k}=2\pi n_{\alpha_0\dots\alpha_k}~,
\eea
where $n_{\alpha_0\dots\alpha_k}$ are constants. Again ${1\over 2\pi}\omega^k$ represents a class in $H^k(M, \bZ)$, iff $n_{\alpha_0\dots\alpha_k}\in \bZ$.

The transition functions of the $\omega^k$ are not unique. Rather they are  specified up to the gauge transformations
\bea
a'^{k-\ell}_{\alpha_0\dots \alpha_{\ell+1}}= a^{k-\ell}_{\alpha_0\dots \alpha_{\ell+1}}+d \zeta^{k-\ell-1}_{\alpha_0\dots \alpha_{\ell+1}}+ \delta \zeta^{k-\ell}_{\alpha_0\dots \alpha_{\ell+1}}~.
\label{egauge}
\eea

To construct $C^{[\omega^k]}_M$ introduce coordinates $y^{k-\ell}_{\alpha_0\dots \alpha_{\ell}}$  and impose the patching conditions
\bea
&&\delta y^{k-\ell}_{\alpha_0\dots \alpha_{\ell+1}}+dy^{k-\ell-1}_{\alpha_0\dots \alpha_{\ell+1}}=a^{k-\ell}_{\alpha_0\dots \alpha_{\ell+1}}~,~~~\ell=2,\dots, k-1~,
\cr
&&\big(\delta y^0_{\alpha_0\dots \alpha_k}-a^0_{\alpha_0\dots \alpha_k}\big)=0 ~\mathrm{mod}~ 2\pi \bZ~,
\eea
where now $y^0$ denote the new angular coordinates.
After acting with $\delta$, it is clear from the last patching condition  that consistency  requires that  $n_{\alpha_0\dots \alpha_{k+1}}\in \bZ$ and so ${1\over 2\pi} \omega^k$ represents a class in $H^k(M, \bZ)$. This is
 the Dirac quantization condition. Note that the construction begins with the introduction of a new coordinate which locally is a $(k-2)$-form as expected from considerations
 that apply to  exceptional field
theories containing a k-form.   Then proceed with the introduction of many new other coordinates at the multiple overlaps of the open sets of the good cover.

The construction of $C^{[\omega^k]}_M$ is independent from the choice of the transition functions in (\ref{egauge})  provided we allow
the new coordinates to transform as
\bea
&&y'^{k-\ell}_{\alpha_0\dots \alpha_{\ell}}=y^{k-\ell}_{\alpha_0\dots \alpha_{\ell}}+\zeta^{k-\ell}_{\alpha_0\dots \alpha_{\ell}}~.
\eea
In addition one can show that $C^{[\omega^k]}_M$ depends only on the class of ${1\over2\pi}\omega^k$ in $H^k(M, \bZ)$.  Furthermore, $C^{[\omega^k]}_M$
obeys the topological geometrization condition. In particular, it is easy to see from the construction above that $dy^{k-2}_\alpha-A^{k-1}_\alpha=dy^{k-2}_\beta-A^{k-1}_\beta$
and so $\pi^*\omega^k=-d (dy^{k-2}-A^{k-1})$ is exact on the C-space.

\subsection{Applications}

Most of the properties and applications we have explored for $C^{[\omega^3]}_M$ can be extended to $C^{[\omega^k]}_M$.  Selectively, on
$C^{[\omega^k]}_M$ one can introduce  an extension
\bea
0\rightarrow \pi^*\Lambda^{k-2}(M)\rightarrow {\cal E}\rightarrow \pi^*TM\rightarrow 0~,
\eea
where  $\Lambda^{k-2}(M)$ is the bundle of (k-2)-forms.
As $dy^{k-2}_\alpha-A^{k-1}_\alpha$ is globally defined on $C^{[\omega^k]}_M$ introduces a splitting of ${\cal E}$ and using this one can introduce a bracket and write a generalized metric in a way similar to that of
$C^{[\omega^3]}_M$ presented in section 3.2.  $C^{[\omega^k]}_M$ provides also a model for a k-gerbe.

In the context of exceptional field theories, the strong section condition, under similar assumptions to the DFT case, will lead to a patching condition
\bea
-\tilde x_\alpha^{k-2}+ \tilde x_\beta^{k-2} = \kappa^{k-2}_{\alpha\beta}~.
\eea
for the (k-2)-form coordinates.  Again this implies that the exceptional spaces
 are  diffeomorphic to $\Lambda^{k-2}(M)$. Such a space cannot satisfy the topological geometrization condition.  Furthermore  if $\kappa^{k-2}_{\alpha\beta}$ are related to the transition
functions of $\omega^k$ at double overlaps with a linear map, then $\omega^k$ represents the trivial class in cohomology. The exceptional spaces are local subspaces of
 $C^{[\omega^k]}_M$  where all coordinates of the latter apart from $x$ and $y^{k-2}$ are set to zero.
These topological considerations lead to the conclusion that for the global definition of exceptional field theories
 many more coordinates are needed in analogy with DFTs.

\section{Whitehead Towers and $C^{[\omega^k]}_M$}

To get a new insight into C-spaces, it is helpful to find alternative constructions
which are not based on local data. For this as a guidance, one can use the topological geometrization
property.  It has been mentioned in \cite{cfolds}, that there is such a construction
 in homotopy theory realized by the Whitehead towers. As we shall see Whitehead towers
 include the Kaluza-Klein construction and provide a homotopy model for spaces that
 satisfy the topological geometrization condition. Moreover when applied to 3-forms have an  intriguing connections to gerbes.

The Whitehead towers  are sequences of fibrations such that
\bea
M\xleftarrow{p_1} X_1\xleftarrow{p_2} X_2\xleftarrow{p_3} X_3\xleftarrow{p_4}\dots
\eea
where the fibre  associated with the $p_n$ projection is the Eilenberg-MacLane space $K(n, \pi_{n-1})$, $\pi_\ell=\pi_\ell(M)$ are the homotopy groups of $M$,
and $X_n$ is n-connected, ie $\pi_\ell(X_n)=0$ for $\ell\leq n$ and also $\pi_{\ell}(X_n)=\pi_{\ell}(M)$ for $\ell>n$. The  Eilenberg-MacLane space $K(m, A)$ has the property
that  $\pi_\ell(K(m, A))=0$ unless $\ell=m$ in which case  $\pi_m(K(m, A))=A$ for any abelian group $A$.

Assuming that $M$ is connected, the description  of $X_1$ begins with the construction of an auxiliary space $Y_1$ which is derived from $M$ after adding cells
to kill all the higher homotopy groups than $\pi_1$. $M$ is included in $Y_1$.  Then a point $z$ is chosen in $Y_1$, and $X_1$ is defined as all paths that begin
at $z$ and end in $M$ as  $M\subset Y_1$. Then $p_1$ is defined as the end point projection of the paths. It turns out that the fibre of this fibration
is homotopic to the loop space $\Omega_*(Y_1)$ which is the fibre over $z$. As by construction $Y_1=K(\pi_1,1)$, one concludes from the homotopy exact sequence of path fibrations that
  $\Omega_*(Y_1)=K(\pi_1, 0)$. As  the only non-vanishing homotopy group is $\pi_0(K(\pi_1, 0))=\pi_1$, from  the homotopy exact sequence of the fibration $M\xleftarrow{p_1} X_1$ one
  finds that $X_1$  homotopic to the universal cover of $M$, ie $X_1$ is simply connected, and $\pi_\ell(X_1)=\pi_\ell(M)$ for all $\ell>1$.  This construction can be repeated
  for $X_1$ to yield $X_2$ and so on, see eg \cite{bott} page 252.

  Next assume that $M$ is simply connected so that we can go straight to the fibration $M\xleftarrow{p_2} X_2$.  The fibre in this case is $K(\pi_2, 1)$
 and $\pi_2=H_2(M,\bZ)=H^2(M, \bZ)$  as $M$ is simply connected. Since $\pi_1(X_2)=\pi_2(X_2)=0$, $H^2(X_2, \bZ)=0$ and so $X_2$ realizes the topological geometrization
  property for $M$ and for all closed 2-forms on $M$. Furthermore, the construction is homotopic to the usual Kaluza-Klein reduction. This is because for $\pi_2=\oplus^m\bZ$,
 the fibre $K(\oplus^m\bZ, 1)$ can be chosen up to a homotopy as $T^m$.  Though there is a difference between $X_2$ and  $C_M^{[\omega^2]}$  as the former
 by construction topologically geometrizes all closed 2-forms while the latter topologically  geometrizes only $\omega^2$. Of course one can repeat the process to construct the C-spaces for all the generators of closed 2-forms in which case both $C_M^{[\omega^2]}$ and $X_2$ will have the same homotopy groups.

  Next let us go one step up the Whitehead tower.  Assume  that $M$ is 2-connected. In such case,  $X_3$ is 3-connected and realizes the topological geometrization property
  for $M$ and for all closed 3-forms on $M$. Furthermore, for $\pi_3=\oplus^m\bZ$, $K(\bZ, 2)=\times^m \bC P^\infty$. This can be easily seen form the homotopy exact sequence of the  Hopf fibration
$S^1\rightarrow S^{2n+1}\rightarrow \bC P^n$ as $n\rightarrow \infty$. In this limit all the homotopy groups of  $S^{2n+1}$ vanish and so the only non-vanishing homotopy group of $\bC P^\infty$ is
 $\pi_2(\bC P^\infty)=\pi_1(S^1)=\bZ$. However $\bC P^\infty$ is also identified as $BU(1)$, the universal classifying space of $S^1$ bundles. Thus
$X_3$ is a fibration over $M$  which arises from gluing a fibre which is the ``space of $S^1$ bundles'' reminiscent of the gerbes according to \cite{ghitchin}.  It is also reminiscent of the emergence of $\bC P^2$ in the exploration of
the topological structure of  $C^{[\omega^3]}_M$.
These raise the question  how  $C^{[\omega^3]}_M$  is related to $X_3$, and whether the former can become a model for the latter. Clearly, the this procedure
for constructing C-spaces using Whitehead towers works for the rest of the cases involving higher degree forms.

For the special case of $M=S^3$ which is of interest in both DFT and theory of gerbes, the Whitehead
fibration $\bC P^\infty \rightarrow X_3\rightarrow S^3$ is a direct generalization of the Kaluza-Klein
construction but now for the 3-form which represents the generator of $H^3(S^3, \bZ)$. In particular observe that $X_3$ and $S^3$ have the same homotopy groups for $n>3$ in direct analogy with the Kaluza-Kelin case where this statement is valid for $n>2$. It is tempting to
assert that $X_3$ provides a homotopy representative for the gerbe.

It should also be noted that  open strings with fixed origin geometrize all form fluxes. To see this, let $P_*(M)$ and $P(M)$ and  be the space of paths in $M$ with and without fixed origin, respectively.
There are two fibrations, $P_*(M)\rightarrow P(M)\rightarrow M$ and $\Omega_*(M)\rightarrow P_*(M)\rightarrow M$. The projection in the first fibration
is defined as the point in $M$ that the path begins while the projection in the second fibration is the path end point projection. $P_*(M)$ is contractible  as every path can be contracted to
the origin. The first fibration implies that the configuration space of open strings is  homotopic to the spacetime. The second fibration geometrizes all form fluxes. Indeed as $P_*(M)$ is contractible, the pull-back of all cohomology classes
of the spacetime to  $P_*(M)$ are cohomogically  trivial.

\section{Concluding remarks}

We have proposed a C-space, $C^{[\omega^3]}_M$, for any  closed 3-form $\omega^k$ on a manifold $M$ which represents a class ${1\over2\pi} [\omega^3] \in H^3(M, \bZ)$. These have been constructed
by introducing appropriate new coordinates and after imposing suitable transition functions which are related to the transition functions of $M$
and the patching data of $\omega^3$ as arise in the \v Cech-de Rham theory. $C^{[\omega^3]}_M$ may not be manifolds. It is confirmed
that $C^{[\omega^3]}_M$ satisfy the topological geometrization condition and provide a geometric explanation for  a generalized geometry structure on the spacetime.
The double spaces of  DFTs are included as local subspaces in $C^{[\omega^3]}_M$. An interpretation of this is that for the global definition
of DFTs additional coordinates are required. We argue that  these new coordinates are necessary on topological grounds
and this should not depend of the details of geometry. However how these can enter in the existing local description of DFTs remains an open problem.

 We have also generalized the construction of C-spaces for any closed k-form on $M$, and we have established that $C^{[\omega^k]}_M$ have similar properties to those of
 $C^{[\omega^3]}_M$. It is expected that these spaces are  required  for the global definition of exceptional field theories.

The construction of $C^{[\omega^3]}_M$ can be done starting from any spacetime with a good cover and a closed 3-form.  As a result
such spaces can be found for all relevant supergravity backgrounds including those of the NS5-branes.  Here we have explored in detail
the 3-torus with a 3-form flux model of \cite{rev3}. We demonstrate how several puzzles associated with the construction of double spaces
for this model \cite{cfolds} are resolved via the use of C-spaces.

Another method to topologically geometrize k-forms in the context of homotopy theory is that of  Whitehead towers. It was emphasized that
for simply connected manifolds and closed 2-forms, the Whitehead is related to the  construction of the usual Kaluza-Klein space $C^{[\omega^2]}_M$ of circle fibrations. This raises the question whether $C^{[\omega^3]}_M$
can be also related to the Whitehead construction for 3-forms and in particular whether the former provide a model for the latter.
Such a relation will elucidate the topological structure of C-spaces.

Although  C-spaces resolve the global patching problem of double spaces, the
additional coordinates   enter linearly in the transition functions and so appear as too special to allow for a full covariance under all required symmetries,
diffeomorphisms and dualities, without any further assumptions on the structure of spacetime. Nevertheless, they may prove to be useful way to proceed.
 In addition, the understanding how  to incorporate
 the additional coordinates in DFT may lead to some new insights into the structure of these theories.

\vskip 1cm

\noindent{\bf Acknowledgements} \vskip 0.1cm
I would like to thank  Jan Gutowski for many helpful discussions. I am partially supported by the STFC grant ST/J002798/1.

\vskip 0.5cm

\setcounter{section}{0}\setcounter{equation}{0}

\setcounter{section}{0}\setcounter{equation}{0}

\end{document}